\magnification=\magstep1
\hsize=125mm
\overfullrule=0pt
\font\title=cmbx10 scaled\magstep2
\baselineskip=6mm

JYFL preprint

\bigskip

\bigskip

\bigskip

\bigskip

\bigskip

\bigskip

\bigskip

\centerline{\title Microscopic Properties of Horizons}

\bigskip

\bigskip

\centerline{Jarmo M\"akel\"a
\footnote{$^*$}{e-mail: jarmo.makela@phys.jyu.fi}}

\medskip

\centerline{\it Department of Physics, University of Jyv\"askyl\"a}

\centerline{\it P.O. Box 35, FIN 40351, Jyv\"askyl\"a, Finland}

\bigskip

\centerline{\bf Abstract}

\medskip

We suggest that all horizons of spacetime, no matter whether they are
black hole, Rindler, or de Sitter horizons, have certain microscopic
properties in common. We propose that these properties may be used as the
starting points, or postulates, of a microscopic theory of gravity.

\vfill\eject

It is a curious historical fact that progress in physics is often made
when a fundamental problem is raised to the status of a postulate.
Something like that was done by Jacobson in 1995.[1] In that time
theoretical physicists were very much puzzled by the result that the
entropy of a black hole is, in natural units, one quarter of its horizon
area, and several explanations, based either on string theory [2]
or canonical quantum gravity [3], were provided some time later. Instead
of attempting
to provide yet another explanation Jacobson assumed that not only black
hole horizon, but also the so called Rindler horizon of an accelerating
observer may be associated with an entropy which is one quarter of its
area. Using this assumption, together with the first law of
thermodynamics, Jacobson was able to {\it derive} Einstein's field
equation describing the interaction between spacetime and the Unruh
radiation observed by an accelerated observer.[4]

In more precise terms, Jacobson's line of reasoning, with slight
modifications of the original idea, may be expressed as follows: Unruh
radiation coming, from the observer's point of view, through the Rindler
horizon, carries energy and momentum which may be stored to the
observer's detector, and the detector becomes heated. As a result,
spacetime in the vicinity of the observer becomes curved and,
consequently, the paths of the light rays determining the Rindler
horizon change. A closer investigation reveals that, in the rest frame of
the observer, the area of the part of the horizon considered by the
observer shrinks during the radiation process. The Unruh radiation with
temperature $T$ obeys the first law of thermodynamics:
$$
\delta Q = T\delta S,\eqno(1)
$$
where $\delta Q$ and $\delta S$, respectively, are the changes of the
heat and the entropy of the detector due to radiation. If one assumes
that between the change $\delta A$ of the horizon area, and the entropy
change $\delta S$ there is the relationship
$$
\delta S = -{1\over 4}\delta A,\eqno(2)
$$
then Eq.(1) gives the relationship between the energy momentum stress
tensor $T_{\mu\nu}$ of the radiation ($\delta Q$ depends on
$T_{\mu\nu}$), and the Ricci tensor $R_{\mu\nu}$ of spacetime ($\delta
A$ depends, through Raychaudhuri equation, on $R_{\mu\nu}$.). Actually,
Eq.(1)
gives precisely Einstein's field equation, which thereby has been
derived by means of purely thermodynamical arguments.

Jacobson's thermodynamical derivation of Einstein's field equation,
based entirely on the first law of thermodynamics, and the assumed
relationship (2) between entropy and horizon area, raises an interesting
question of a possibility that maybe Einstein's general theory of
relativity describes just thermodynamics of spacetime and matter
fields. In that case the spacetime metric $g_{\mu\nu}$ is probably not a
fundamental field of nature, and all attempts to quantize Einstein's
equation canonically would be, to quote Jacobson's words, "no more
approriate than it would be to quantize the wave equation for sound in
air".[1]

   The thermodynamical properties of any system follow from the
statistical mechanics of that system which, in turn, follows from its
microscopic properties. It would be very interesting to find the
physical laws governing the microscopic properties, and hence the 
statistical  mechanics, of spacetime, but if Jacobson's  
provocative statement is true, a straightforward application of the
rules of quantum mechanics to Einstein's field equation (canonical
quantization, for instance) is of no help. At the present state of
research we must just postulate those laws. The laws must be postulated
in such a way that, among other things, Einstein's field equation is
produced in the thermodynamical limit. For the sake of simplicity we
shall assume that spacetime, at least effectively, is described by a
four-dimensional (pseudo) Riemannian manifold. An advantage of this
assumption is that the whole tensor machinery of Riemannian geometry is
still in our service.

  The crucial step in Jacobson's derivation of Einstein's field equation
was the relationship (2) between horizon area and entropy, and our task
is to find the simplest possible postulates which imply that
relationship. During the past thirty years or so Bekenstein and others
have produced an enormous amount of evidence supporting the
proposal that the area of the event horizon of a black hole has an equal 
spacing in its spectrum.[5] Our first postulate therefore reads:

\vfill\eject

{\bf Postulate 1:} {\it When one measures the area of a horizon, or any
part of it, the possible outcomes of measurements are:}
$$
A_n = n\cdot A_0,\eqno(3)
$$
{\it where} $n$ {\it is a non-negative integer, and} $A_0$ {\it is a
constant}.

\bigskip

Now, the observer may measure the whole area of the horizon by dividing
the horizon into parts, measuring the area of each individual part, and
finally adding together the results of measurements. For the horizon
area $A=nA_0$ the maximum number of parts is $n$. It is a nice  exercise
of combinatorics to show that the number of different ordered $p$-tuples
$$
(m_1,m_2,...,m_p)
$$
such that $1\leq p\leq n$, $m_j\in {\cal
Z}^+\,\,\forall\,j=1,2,3,...,p$, and 
$$
m_1 + m_2 + m_3 + ... + m_p = n,\eqno(4)
$$
is
$$
\Omega(n) = 2^{n-1}.\eqno(5)
$$
In other words, the areas of the individual parts of the horizon may be
summed over to $nA_0$ in $2^{n-1}$ ways. Each ordered $p$-tuple
represents a certain combination of the areas of the parts of the horizon
with fixed total area. We identify each such combination as a microstate
of the
horizon. Hence we get the following postulate:

\bigskip

{\bf Postulate 2:} {\it The number of microstates corresponding to the
same macrostate of the horizon is equal to the number of different
combinations of the areas of its parts.}

\bigskip

Our Postulates 1 and 2 imply that the horizon has entropy $S_h$ which,
for macroscopic horizons, is proportional to the area:
$$
S_h = k_B\ln\Omega(n) = (n-1)k_B\ln 2 \approx {{k_B\ln
2}\over{A_0}}A_n,\eqno(6)
$$
where $k_B$ is Boltzmann's constant. For horizons having infinite area,
such as the Rindler horizon, this entropy may be associated with the
considered finite part of the horizon. In the process we have introduced
the area $A_0$ which may be viewed, in our approach, as a fundamental
constant of nature. The requirement that the entropy of the horizon is,
in natural units, one quarter of its area, gives the following
relationship between Newton's gravitational constant $G$ and the area
$A_0$:
$$
G = {{A_0c^3}\over{4\hbar\ln 2}},\eqno(7)
$$
and therefore
$$
A_0 \approx 7.23\times  10^{-70}m^2.\eqno(8)
$$

    Although our Postulates 1 and 2 imply that the entropy of the
horizon is proportional to its area, they say nothing about the entropy
of the radiation emitted by the horizon. Therefore we state:

\bigskip

{\bf Postulate 3:} {\it In thermal equilibrium the sum of the entropies
of the horizon and the radiation is constant.}

\bigskip

In other words, the entropy of the horizon decreases exactly as much as
the entropy of the radiation increases. As a whole, our Postulates 1, 2
and 3 imply the relationship (2) between the horizon area, and the
entropy of radiation.

    There is only one part missing from our set of postulates. To be able to 
derive Einstein's field equation from the first law of thermodynamics of
Eq.(1) we need a postulate which tells that the radiation emitted by the
horizon has a certain temperature. Since entropy is proportional to
area, and between energy and entropy there is the relationship given by
Eq.(1), we need a postulate which tells the relationship between the area
of the horizon, and the amount of energy which may be extracted out from
the horizon.

      As it is well known, the temperature of the radiation emitted by
any horizon is, in SI units, [4]
$$
T = {{\hbar\kappa}\over{2\pi k_Bc}},\eqno(9)
$$
where $\kappa$ is the surface gravity of the horizon. Hence, between
the energy change $dE$ and the change $dA$ of the area of the horizon
there is the following relationship [6]:
$$
dE = {{\hbar\ln 2}\over{2\pi A_0c}}\kappa\,dA.\eqno(10)
$$
If one assumes that the horizon is a Rindler horizon one finds, 
using a similar chain of reasoning as Jacobson in his paper, that 
Eq.(10) implies Einstein's field equation for any radiation coming, 
from the accelerated observer's point of view, from the horizon. In 
other words, Eq.(10), when written for an arbitrary Rindler horizon,
is just another expression for Einstein's field equation when the matter 
part consists purely of radiation. It can be written in the integral form:
$$
\int_{E_i}^{E_f}{{dE}\over{\kappa(E)}} =  {{\hbar\ln 2}\over{2\pi
A_0c}}(A_f - A_i),\eqno(11)
$$
where $E_f$ and $E_i$, respectively, are the amounts of energy which may be 
extracted out from the horizon in the initial state $i$ and the final state $f$,
and $A_i$ and $A_f$ are the corresponding horizon areas. However, Postulate 1 
states that the area of the horizon has a discrete spectrum with equal spacing. 
Because of that, we write our last postulate, a sort of "quantized Einstein 
equation", in the form:

\bigskip

{\bf Postulate 4:} {\it If} $E_i$ {\it and} $E_f$ {\it are the initial and
the final energies which may be extracted from the horizon, and} $n_i A_0$
{\it and} $n_f A_0$ {\it are the corresponding horizon areas, then}
$$
\int_{E_i}^{E_f} {{dE}\over{\kappa(E)}} = {{\hbar\ln 2}\over{2\pi c}}(n_f
- n_i),\eqno(12)
$$
{\it where} $\kappa(E)$ {\it is the surface gravity of the horizon.}

\bigskip

   Among other things, this postulate quantizes the masses of black
holes, and the energies of the quanta of emitted radiation.[7] Of course,
the energies given by Eq.(12) must be red shifted, in curved spacetime and
for accelerated observers, by the factor $(-g_{00})^{-1\slash 2}$. 

   So far it may have remained somewhat obscure why we chose to identify,
in
Postulate 2, the microstates of the horizon as the different combinations
of the areas of its parts. In the light of Postulates  3 and 4, however,
everything becomes clear: Postulate 4 implies that the energy of the
emitted quantum of radiation depends on the accompanied change of the
horizon area which, in turn, is always an integer times the fundamental
area $A_0$. Because the total area of the horizon is also a certain
integer $n$ times the fundamental area $A_0$, it follows that the number
of different combinations of the energies of the quanta of radiation is
equal to the number --which we found to be $2^{n-1}$-- of the different
combinations of the areas of the parts of the horizon. In other words,
the radiation may come out from the horizon in $2^{n-1}$ ways. Because of
that, the maximum amount of entropy carried by the radiation out from the
horizon is $k_B\ln(2^{n-1})$. According to Postulate 3, however, the
entropy of the horizon decreases exactly as much as the entropy of the
radiation increases. Therefore the maximum entropy of the horizon, too,
is $k_B\ln(2^{n-1})$. This may well provide the simplest conceivable
explanation to the black hole entropy.

     It should be noted that our postulates are assumed to be valid for
any horizon, no
matter whether that horizon is a black hole horizon, a de Sitter horizon,
or a Rindler horizon. (If the horizon is infinite, our postulates are
valid for any of its finite parts.) The postulates are in agreement with
everything
we know about the properties of horizons. Usually, the physical results
concerning the horizons 
involved in our postulates are derived by means of arguments based on
Einstein's classical general relativity, and on the general principles of
quantum mechanics and thermodynamics. The idea of this paper, however, has
been to suggest that perhaps this line of reasoning should be turned
upside down: Instead of trying to obtain the results from general
relativity, we take these results to be the starting points, or
postulates, of a microscopic theory of gravitation. Indeed, our extremely
general and absurdly simple postulates concerning the microscopic
properties of the horizons of spacetime imply, by means of Jacobson's
reasoning, Einstein's general relativity in the classical limit, and they
predict, among other things, the Hawking and the Unruh effects, together
with the result that the entropy of a black hole is one quarter of its
horizon area. No doubt, our postulates might have a certain taste of being
rather {\it ad hoc}, nor do they say anything about the microscopic
structure of spacetime. Because of that they should certainly not be
expected to be anywhere near the very fundamental, underlying postulates
of quantum gravity. Nevertheless, one is perhaps not entirely able to
avoid the feeling that, at the present state of research, our postuölates
satisfy many of the requirements one may reasonably pose for the
postulates of a microscopic theory of gravity. It will be interesting to
see whether a straightforward application of these postulates will predict
new, so far unimagined, phenomena of nature.

\bigskip

\centerline{\bf Acknowledgments}

\bigskip

I am grateful to Markku Lehto, Jorma Louko and Markus Luomajoki for their
constructive criticism and useful discussions during the preparation of
this paper.

\bigskip

\centerline{\bf Notes and References}

\bigskip

[1] T. Jacobson, Phys. Rev. Lett. {\bf 75} 1260 (1995)

\medskip

[2] See, for example, A. Strominger and C. Vafa, Phys. Lett. {\bf B379},
99 (1996).

\medskip

[3] See, for example, A. Ashtekar, A. Baez, J. Korichi and K. Krasnov,
Phys. Rev. Lett. {\bf 80}, 904 (1998).

\medskip

[4] See, for example, R. M. Wald, {\it Quantum Field
Theory in Curved Spacetime and Black Hole Thermodynamics} (The University
of Chicago Press, Chicago and London, 1994), and W. G. Unruh, Phys. Rev.
{\bf D14}, 870 (1976).

\medskip

[5] J. D Bekenstein, Lett. Nuovo Cimento {\bf 11}, 467 (1974). For an
updated review and a list of references on this subject see,
for example, J. D. Bekenstein, "The Case for Discrete Energy Levels of a
Black Hole", hep-th/0107045. A comprehensive list of references may also
be
found in J. M\"akel\"a, P. Repo, M. Luomajoki and J. Piilonen, Phys. Rev.
{\bf D64}, 0240018 (2001).

\medskip

[6] To find the actually observed temperature of radiation and the energy
change of the horizon, the quantity $T$ of Eq.(9) as well as the quantity
$dE$ of Eq.(10) must be multiplied by an appropriate red shift factor. For
an accelerated observer, for instance, this factor is, in natural unis,
the proper acceleration of the observer.

\medskip

[7] Discrete mass eigenvalues and the thermal spectrum of black hole
radiation do not necessarily contradict with each other. (See, for
example, J. M\"akel\"a, Phys. Lett. {\bf B390}, 115 (1997).)

\bye